\documentclass[twocolumn,pra,showpacs,superscriptaddress]{revtex4}
\usepackage{amsmath}
\usepackage{graphicx}

\input{tcilatex}

\begin{document}

\title{Molecular Ensemble Based Remote Quantum Storage \\
for Charge Qubit via Quasi-Dark State}
\author{H. R. Zhang }
\affiliation{Institute of Theoretical Physics, Chinese Academy of Sciences, Beijing,
100080, China}
\author{Y. B. Gao}
\affiliation{College of Applied Sciences, Beijing University of Technology, Beijing,
100022, China}
\author{Z. R. Gong}
\affiliation{Institute of Theoretical Physics, Chinese Academy of Sciences, Beijing,
100080, China}
\author{C. P. Sun}
\affiliation{Institute of Theoretical Physics, Chinese Academy of Sciences, Beijing,
100080, China}
\date{\today}

\begin{abstract}
We propose a quantum storage scheme independent of the current time-control
schemes, and study a ``quantum data bus'' (transmission line resonator) in a
hybrid system consisting of a circuit QED system integrated with a cold
molecular ensemble. Here, an effective interaction between charge qubit and
molecule is mediated by the off-resonate field in the data bus.
Correspondingly, the charge state can be mapped into the collective
quasi-spin state of the molecular ensemble via the standard dark state based
adiabatic manipulation.
\end{abstract}

\maketitle

\section{introduction}

Quantum storage is crucial to quantum information processing~\cite%
{Knill,Zhou,Barenco}. Its importance rests with two issues: mapping the
qubits with shorter coherence time into a system with relatively longer
coherence time~\cite{Pazy,Song,Taylor}; and storing the quantum states of
the physical system difficultly operated into the states of the system
manipulated feasibly~\cite{Sunm}. Superconducting qubits, such as the charge
qubits, flux qubits and phase qubits, are hopeful candidates for the basic
blocks in the architecture of quantum computer~\cite{Makh} since they
exhibit feasibility to be well manipulated and well integrated~\cite%
{charge,Vion,flux,Mart,Yu,Pash}. Currently the decoherence of the
superconducting qubits is relatively fast so that no enough logical
operations can be performed for a practical quantum computing. Thus people
expect that the superconducting qubits can be stored in some system for
which the decoherence time is long enough so that the stored qubits can be
operated within an enough time.

To overcome the disadvantage of superconducting qubits a hybrid quantum
processor has been suggested interfacing the solid state system and the
molecular ensemble~\cite{holo,hybi}, in which the ensemble of cold polar
molecules was used as a long-lived quantum memory storing the information of
superconducting quantum qubits by collective spin state via microwave Raman
processes in a high-Q stripline cavity~\cite{hybi}. Further more, a
holographic quantum computing has been demonstrated using this protocol~\cite%
{holo}. These conceptual designs used the real photons as the data bus, and
the corresponding schemes depend on an exact time control to switch
photon-qubit and photon-molecular couplings effectively on and off. Based on
the above considerations, to avoid the difficulty in exactly controlling
evolution time , we study the data bus role of quantum transmission line. By
adiabatically eliminating the variables of the transmission line as quantum
data bus, we obtain dark state for the effective coupling between molecular
ensemble and the charge qubit. And then adiabatically manipulating the
control field, the superconducting charge qubit can be mapped into the
collective quasi-spin state of the molecular ensemble, and a reversing
operation in a latter time will regenerate the charge qubit state. During
the adiabatic passage the transmission line is virtual excited and no real
photons in the transmission line are exchanged between the charge qubit and
the molecular ensemble. In this sense the effective dark state of the charge
qubit and the molecular ensemble computing system plays crucial role.

The paper is organized as follows. In Sec.~II we describe the circuit QED
setup hybridized with embedded molecular ensemble. In Sec.~III, we
demonstrate how the off-resonate coplanar cavity field induces the effective
coupling of charge qubit to collective excitation of molecular ensemble and
the existence of the quasi-dark state at parameter resonance. In Sec.~IV, we
show that using the effective dark state, quantum storage of the charge
qubit state can be implemented through an adiabatic manipulation.

\section{Modeling the hybrid system in terms of collective excitations}

\begin{figure}[tph]
\includegraphics[bb=15 490 376 809, width=6.5cm]{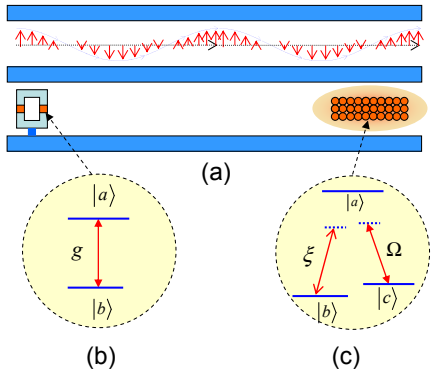}
\caption{Schematic diagram of a hybrid system where the cold
molecular ensemble is coupled to a circuit QED system consisting of
a Cooper-pair box and a stripline cavity in (a). The Cooper-pair box
(as a two level system (b)) and cold molecular ensemble are at the
adjacent antinodes in the stripline cavity. For the molecular
ensemble, each molecule (b) has three
energy levels where the level $|a\rangle $ is coupled to $|b\rangle $ and $%
|c\rangle $ by a quantum field in stripline cavity and a classical control
field, respectively. }
\label{setup}
\end{figure}

We consider a hybrid system shown in Fig.~\ref{setup} that consists of a
cold molecular ensemble, a Cooper-pair box (charge qubit) and the stripline
cavity formed by a transmission line resonator~\cite{holo,Tordrup08}. The
molecular ensemble consists of $N$ identical and noninteracting $\Lambda $%
-type molecules, each with a ground state $|b\rangle $, an excited state $%
|a\rangle $ and a metastable lower excited state $|c\rangle $ (see in Fig.~%
\ref{setup} (b)). At the antinodes of quantum electromagnetic field in
stripline cavity, the polar molecules with large electric-dipole moment
achieve the strong coupling, similarly as that in the circuit QED system~%
\cite{Wallraff04}.
We place the Cooper-pair box and the molecular ensemble in two antinodes at
a distance. In some proper conditions~\cite{Blais04}, the Cooper-pair box
can be reduced to a two-level quantum system (charge qubit) with
Jaynes-Cummings type coupling to the single mode cavity field. The
Hamiltonian of the circuit QED system reads%
\begin{equation}
H_{1}=\frac{\omega _{g}}{2}\sigma _{z}+\omega a^{\dag }a+g(a\sigma
_{+}+h.c.),
\end{equation}%
where $a$ ($a^{\dag }$) is the quantum cavity field annihilation and
(creation) operator, $\sigma _{-}$ ($\sigma _{+}$) the lowering (raising)
operator, and $g$ the strength of the interaction between the stripling
cavity field and the Cooper-pair box. The parameters $\omega _{g}$ and $%
\omega $ denote the frequency of charge qubit and the quantum cavity field
respectively. Here we have assumed $\hbar =1$ for convenience in this paper.
Initially we can prepare the polar molecule in the ground state $|b\rangle $%
, which is coupled to the excited state $|a\rangle $ by the quantum
electromagnetic field in the stripline cavity. While the coupling of the
ground state $|b\rangle $ to the metastable excited state $\left\vert
c\right\rangle $ is realized through the classical control field with the
Rabi frequency $\Omega $. Thus the Hamiltonian for the coupled molecular
ensemble-stripline cavity can be written as
\begin{eqnarray}
H_{2} &=&\sum_{j=1}^{N}(\omega _{a}\left\vert a\right\rangle
_{jj}\left\langle a\right\vert +\omega _{c}\left\vert c\right\rangle
_{jj}\left\langle c\right\vert )+\omega a^{\dag }a  \notag \\
&&+\sum_{j=1}^{N}\left( \xi a\left\vert a\right\rangle _{jj}\left\langle
b\right\vert +\Omega e^{-i\omega _{f}t}\left\vert a\right\rangle
_{jj}\left\langle c\right\vert +h.c.\right),
\end{eqnarray}%
where $\xi $ is the interaction strength of the quantum cavity field to the
molecular ensemble, $\omega _{a}$ and $\omega _{c}$ the molecule's energy
level spacings with respect to the excited state and the metastable excited
state respectively, and $\omega _{f}$ the frequency of the control field.

Here we adopt the quasi-spin wave approach developed in Ref.~\cite{Sunm} and
invoke the bosonic operators
\begin{equation}
A=\frac{1}{\sqrt{N}}\sum_{j=1}^{N}|b\rangle _{jj}\langle a|,\qquad C=\frac{1%
}{\sqrt{N}}\sum_{j=1}^{N}|b\rangle _{jj}\langle c|
\end{equation}%
to describe the collective excitation of the ensemble. To this end we
rewrite the above Hamiltonian $H=H_{0}+H_{I}$ in terms of the free part
\begin{equation}
H_{0}=\frac{\omega _{g}}{2}\sigma _{z}+\omega a^{\dag }a+\omega _{a}A^{\dag
}A+\omega _{c}C^{\dag }C
\end{equation}%
and the interaction part
\begin{eqnarray}
H_{I} &=&g(a\sigma _{+}+h.c.)+\zeta \left( aA^{\dag }+h.c.\right)  \notag \\
&&+\Omega \left( e^{-i\omega _{f}t}A^{\dag }C+h.c.\right) ,
\end{eqnarray}
where $\zeta =\xi \sqrt{N}$. In the large $N$ limit with the low excitation
condition that there are only a few atoms occupying the level $|a\rangle $
or $|c\rangle $~\cite{Liu01}, the operator $A$ ($C$) behaves as a boson
which satisfies the bosonic commutation relation
\begin{equation}
\left[ A,A^{\dag }\right] =1,\left[ C,C^{\dag }\right] =1.
\end{equation}%
In the same limit, we also have
\begin{equation}
\left[ A,C\right] =0,\left[ A,C^{\dag }\right] \rightarrow 0,
\end{equation}%
which means that two boson modes described by $A$ and $C$ are independent
with each other.

\section{Induced effective coupling of charge qubit to collective
excitations of molecular ensemble}

We emphasize again that, the current hybrid system based quantum storage
protocol requires an exact time control for couplings of charge qubit to
photons and photons to molecular ensemble. Namely, one can switch on (off)
the couplings exactly in an arbitrarily given instance. Such a time control
process includes two steps: a) couple the photon to the charge qubit
firstly; b) then switch off the previous coupling and switch on the coupling
of the photon to the molecular ensemble again. Due to the \textquotedblleft
time-energy uncertainty\textquotedblright , this kind of time control
protocol would inevitably result in experimental errors. It is much
appreciated that we can avoid these errors. With these considerations, we
detune stripline cavity field slightly from both the charge qubit and the
molecular ensemble. Thus the effective interaction between the charge qubit
and the molecular ensemble will be induced by adiabatically eliminating
photon degrees of freedom.

\begin{figure}[th]
\includegraphics[bb=11 560 322 791, width=5.5cm]{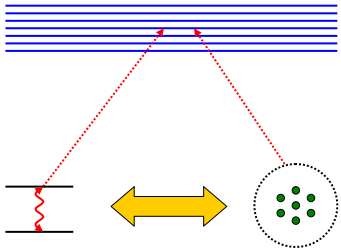}
\caption{Effective interaction between the charge qubit and the molecular
ensemble induced by the virtual photon exchange. }
\label{ske}
\end{figure}

Initially, we prepare the microwave field in the vacuum state, then the
interaction between the charge qubit and the molecular ensemble is induced
by the virtual excitation of the transmission line field, with no exchange
of the real photons between the charge qubit (molecular ensemble) and the
field. The process can be schematically shown in Fig.~\ref{ske}. Here both
the qubit and the molecular ensemble interact simultaneously with the
continuum of the cavity field whose spectral structure is illustrated in the
top of Fig.~\ref{ske}. We will show as follows that, under some conditions
related to the large detuning the virtual photon transition happens between
the qubit (molecular ensemble) and the transmission line cavity. The
photon-assisted transitions can not happen in practice, but the effective
interaction between qubit and the molecular ensemble is induced, which is
similar to the conventional contact interaction. Therefore, our present
protocol does not need an exact time control for the coupling of the photon
qubits to the charge qubit (to the molecular ensemble).
Next we study in details the virtual photon exchange process by deriving the
effective Hamiltonian through a canonical transformation. In the rotating
frame in terms of a time dependent unitary operator $U=\exp (i\omega
_{f}tC^{\dag }C)$, we get a time independent Hamiltonian
\begin{equation}
H^{R}=U^{\dag }HU-iU^{\dag }\partial _{t}U.  \label{ham}
\end{equation}%
When we adopt the frequency-matching condition, $\omega _{a}=\omega
_{c}+\omega _{f}$, i.e., resonantly coupling the metastable lower excited
state $|c\rangle $ to the excited state $|a\rangle $, the time-independent
Hamiltonian (\ref{ham}) can be decomposed into tow parts, $H^{R}=H_{0}+H_{I}$%
, including the \textquotedblleft free\textquotedblright\ part Hamiltonian%
\begin{equation}
H_{0}=\frac{\omega _{g}}{2}\sigma _{z}+\omega a^{\dag }a+\omega _{a}A^{\dag
}A+\omega _{a}C^{\dag }C+\Omega (A^{\dag }C+h.c.)
\end{equation}%
and the interaction part Hamiltonian%
\begin{equation}
H_{I}=g(a\sigma _{+}+h.c.)+\zeta (aA^{\dag }+h.c.),
\end{equation}%
which describes the coupling of stripline cavity field to both the charge
qubit and the molecular ensemble.

Using Fr\"{o}hlich's transformation~\cite{froli,froli1,froli2}, we can
obtain the effective Hamiltonian
\begin{equation}
H_{eff}=e^{-S}H^{R}e^{S}\cong H_{0}+\frac{1}{2}[H_{I},S],  \label{effham}
\end{equation}%
where $S$ is an anti-Hermitian operator satisfying
\begin{equation}
H_{I}+[H_{0},S]=0.  \label{condi}
\end{equation}%
The corresponding anti-Hermitian operator $S$ for Fr\"{o}hlich
transformation adopts the following form
\begin{equation}
S=(\eta _{1}a\sigma _{+}+\eta _{2}aA^{\dag }+\eta _{3}aC^{\dag })-h.c,
\label{hams}
\end{equation}%
where the coefficients $\eta _{1}$, $\eta _{2},$ and $\eta _{3}$ will be
determined by Eq.~(\ref{condi})
\begin{eqnarray}
\eta _{1} &=&\frac{g}{\omega -\omega _{g}},  \notag \\
\eta _{2} &=&\frac{\zeta \left( \omega -\omega _{a}\right) }{\left( \omega
-\omega _{a}\right) ^{2}-\Omega ^{2}},  \notag \\
\eta _{3} &=&\frac{\zeta \Omega }{\left( \omega -\omega _{a}\right)
^{2}-\Omega ^{2}}.
\end{eqnarray}

In order to obtain the effective coupling between the charge qubit and the
molecular ensemble we consider the lightly off resonant case and take
appropriate experimental conditions so that the coefficients $\eta _{1}$, $%
\eta _{2}$ and $\eta _{3}$ are all as small as enough, i.e., $|\eta _{1}|$, $%
|\eta _{2}|$, $|\eta _{3}|\ll 1$, which are also requisite by the usual
second order perturbation theory. We assume that the microwave cavity field
is initially in the vacuum state, i.e., $\left\langle a^{\dag
}a\right\rangle =0$. Averaged over the vacuum state of the photon field, in
the two photon resonance case, the effective Hamiltonian~(\ref{effham}) in
the interaction picture reads,
\begin{equation}
H_{int}=\Delta B^{\dag }B+(g_{m}\sigma _{-}B^{\dag }+\Omega _{d}B^{\dag
}D+h.c.),  \label{hme}
\end{equation}%
where the bosonic operators $B$ and $D$ are defined\
\begin{equation}
B=\alpha A+\beta C\text{, }D=\beta A-\alpha C
\end{equation}%
with some time-dependent coefficients
\begin{equation}
\alpha =\frac{\left( \eta _{2}g+\eta _{1}\zeta \right) }{\sqrt{(\eta
_{3}g)^{2}+\left( \eta _{2}g+\eta _{1}\zeta \right) ^{2}}},  \label{alph}
\end{equation}%
\begin{equation}
\beta =\frac{\left( \eta _{3}g\right) }{\sqrt{(\eta _{3}g)^{2}+\left( \eta
_{2}g+\eta _{1}\zeta \right) ^{2}}}.  \label{beta}
\end{equation}%
The effective Hamiltonian~(\ref{hme}) shows that the photon degrees of
freedom is effectively eliminated and it mediates an effective interaction
between the charge qubit and the molecular ensemble. The two photon
resonance condition is
\begin{equation}
\omega _{B}-\omega _{g}^{\prime }=\omega _{B}-\omega _{D}=\Delta
\label{reson}
\end{equation}%
with $\omega _{B}$ and $\omega _{D}$ being the effective energy level
spacings with respect to the states of $B$-excitation and $D$-excitation
respectively, $\omega _{g}^{\prime }$ the effective frequency of the charge
qubit after the Fr\"{o}hlich transform. The complicated coefficients in $%
H_{int}$ are listed as follows
\begin{equation}
g_{m}=-\frac{1}{2}\sqrt{(\eta _{3}g)^{2}+\left( \eta _{2}g+\eta _{1}\zeta
\right) ^{2}},  \label{coe}
\end{equation}%
\begin{equation}
\omega _{g}^{\prime }=\omega _{g}-\eta _{1}g,  \label{coe1}
\end{equation}%
\begin{equation}
\omega _{B}=\omega _{a}-\eta _{2}\zeta \alpha ^{2}+2\left( \Omega -\frac{1}{2%
}\eta _{3}\zeta \right) \alpha \beta ,  \label{coe2}
\end{equation}%
\begin{equation}
\omega _{D}=\omega _{a}-\eta _{2}\zeta \beta ^{2}-2\left( \Omega -\frac{1}{2}%
\eta _{3}\zeta \right) \alpha \beta ,  \label{coe3}
\end{equation}%
\begin{equation}
\Omega _{d}=-\eta _{2}\zeta \alpha \beta +\left( \Omega -\frac{1}{2}\eta
_{3}\zeta \right) (\beta ^{2}-\alpha ^{2}).  \label{coef}
\end{equation}

In order to find invariant subspaces of $H_{int}$, we define the dark state
operator
\begin{equation}
F=\sigma _{-}\cos \theta -D\sin \theta
\end{equation}%
with $\theta (t)$ satisfying $\tan \theta (t)=g_{m}/\Omega _{d}(t)$. It
mixes the states of the qubit excitation and the collective molecular
excitations. We introduce the composite vacuum state $|\mathbf{0}\rangle =$\
$|\mathbf{b}\rangle \otimes |g\rangle $, where $|\mathbf{b}\rangle $ is the
ground state with all $N$ molecules staying in the same single particle
ground state $|b\rangle$~\cite{Sunm}. Evidently an eigenstate of $H_{int}$
with vanishing eigenvalue is constructed as
\begin{equation}
|\Psi \rangle =F^{\dag }|\mathbf{0}\rangle.  \label{dark}
\end{equation}%
We notice that $|\Psi \rangle $ is not a ideal dark state, actually it is a
superposition state of the $A$-excitation concerning the excited state which
will decay during the usual manipulation process. In this sense, we name it
quasi-dark state.

To construct other eigenstates of the interaction Hamiltonian, we introduce
the operators as follows
\begin{equation}
Q_{\pm }=\sqrt{\frac{\Theta \pm \Delta }{2\Theta }}B\pm \sqrt{\frac{\Theta
\mp \Delta }{2\Theta }}(\sigma _{-}\sin \theta +D\cos \theta ),
\end{equation}%
where $\Theta =\sqrt{\Delta ^{2}+4(g_{m}^{2}+\Omega _{d}^{2})}.$ Using these
operators, we can construct the eigenstates $|\Phi _{\pm }\rangle =Q_{\pm
}^{\dag }|\mathbf{0}\rangle $, with the corresponding eigenvalues $E_{\pm
}=\pm \sqrt{(\Theta \pm \Delta )(g_{m}^{2}+\Omega _{d}^{2})/(\Theta \mp
\Delta )}$.

\section{Quasi-Dark state based adiabatic manipulation for quantum storage}

In this section, we will show how to store the quantum information of the
charge qubit in the molecular ensemble by using the above quasi-dark state.
In this quantum storage scheme, adiabatically modulating the control field,
the charge qubit state can be mapped into the collective excitation state of
the molecular ensemble in appropriate experimental conditions.

Now we consider whether the charge qubit state can be \ mapped into the
molecular ensemble effectively. We assume that the Rabi frequency of the
control field is much smaller than the frequency detuning between the cavity
field and the $A$-excitation of molecule, i.e., $|\Omega |\ll |\omega
-\omega _{a}|$. Then the values of $\alpha $ and $\beta $ are approximated
as $\alpha $ $\approx 1$ and $\beta \approx 0$ from the Eq.~(\ref{alph})
and~(\ref{beta}). In this case, there exist other eigenstates with non-zero
eigenvalues besides the quasi-dark state. To avoid the system to transit to
other eingenstates and keep itself within the quasi dark state, the
manipulation of the control field should satisfy the adiabatic condition
\begin{equation}
|\frac{\langle \Phi _{\pm }|\partial t|\Psi \rangle }{E_{\pm }-0}|=\frac{%
g_{m}(\Theta +|\Delta |)}{\sqrt{\Theta (\Theta -|\Delta |)(g_{m}+\Omega
_{d})^{3}}}|\dot{\Omega}|\ll 1
\end{equation}%
according the quantum adiabatic theorem \cite{adi1,adi2}. From the Eqs.~(\ref%
{coe}--\ref{coef}) of the coefficients, with the adiabatic decrease of the
control field, the mixing angle will increase from $0$ to $\pi /2$.

\begin{figure}[th]
\includegraphics[bb=15 626 460 808, width=8cm]{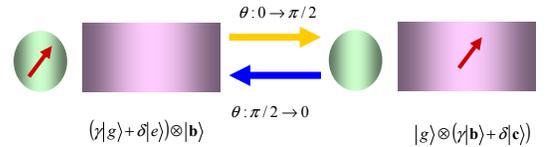}
\caption{Mapping the charge qubit into the molecular ensemble by virtual
photon transition. }
\label{ske2}
\end{figure}

Because the composite vacuum state $|g\rangle \otimes |\mathbf{b}\rangle $
is an eigenstate of the interaction Hamiltonian also, if the mixing angle $%
\theta $ changes from $0$ to $\pi /2$, an arbitrary superposition of charge
qubit state will undergo the following evolution through the quasi-dark
state,
\begin{equation}
(\gamma |g\rangle +\delta |e\rangle )\otimes |\mathbf{b}\rangle
\longrightarrow |g\rangle \otimes (\gamma |\mathbf{b}\rangle +\delta (\alpha
|\mathbf{c}\rangle +\beta |\mathbf{a}\rangle )),
\end{equation}%
where $|\mathbf{a}\rangle =A^{\dag }|\mathbf{b}\rangle $, $|\mathbf{c}%
\rangle =C^{\dag }|\mathbf{b}\rangle $. This means that we can transfer the
information of the charge qubit state into the collective excitation state
of the molecular ensemble without any real photon exchange. This storage
process can be schematically shown in Fig.~\ref{ske2}. Here, rotating the
mixing angle $\theta $ from $0$ to $\pi /2$ by changing the strength of the
control field adiabatically, the information of the charge qubit state is
stored in the molecular ensemble. And the reverse manipulation will make the
molecular ensemble release the information to the charge qubit again.

We have approximately considered the approach of mapping the charge qubit
into the molecular ensemble. To realize the storage of a charge qubit, next
we will consider some experimental data and show how the coefficients change
with respect of the control field in details.

\begin{figure}[th]
\includegraphics[]{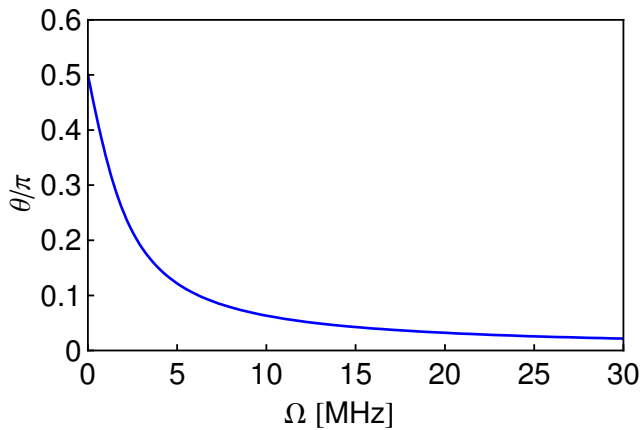}
\caption{Variance of the mixing angle $\protect\theta$ by adiabatic
modulation of the control field with the Rabi frequency $\Omega$ changing
from $0$ to $30 \text{MHz}$. }
\label{thetav}
\end{figure}

\begin{figure}[th]
\includegraphics[]{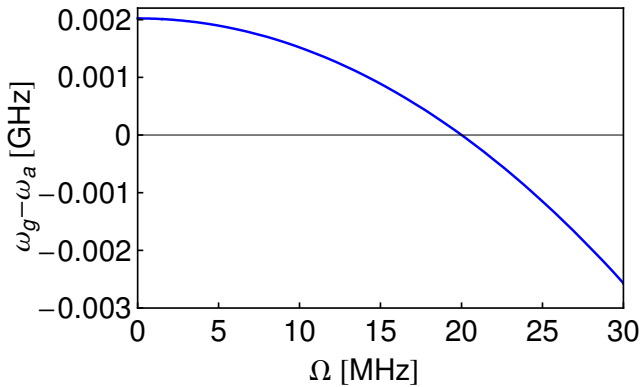}
\caption{Modulation of the frequency of the charge qubit with the
change of the control field. Here the variance of the qubit
frequency is denoted by the frequency difference between charge
qubit and the $A$-excitation of the molecular ensemble and the unit
is GHz.} \label{omegag}
\end{figure}

\begin{figure}[th]
\includegraphics[]{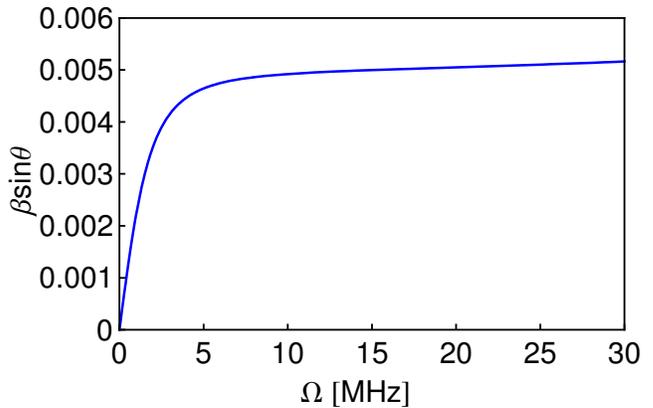}
\caption{Variance for the coefficient of the excited state during the
storage process. When the Rabi frequency of the control field adiabatically
varies from $0$ to $30\text{MHz}$, the magnitude (from 0 to 0.005) of the
coefficient $\protect\beta\sin\protect\theta$ of the excited state is very
small.}
\label{betat}
\end{figure}

From the references~\cite{Wallraff04,hybi}, we choose the experimental data:
$g=20\text{MHz}$, $\zeta =20\text{MHz}$, $\omega =6.044\text{GHz}$ and $%
\omega _{a}=5.844\text{GHz}$. The effective coupling $g_{m}=2.01\text{MHz}$
which almost does not depend on the control field. The mixing angle $\theta $
changes from $\pi /2$ to $0$ by adiabatically modulating the control field
with the Rabi frequency $\Omega$ varying from $0$ to $30\text{MHz}$, (see,
e.g., Fig.~\ref{thetav}). The frequency of charge qubit needs to be
modulated (see, e.g., Fig.~\ref{omegag}) with the variance of the control
field to satisfy the resonance condition (\ref{reson}). The change range is
about $5\text{MHz}$, which is very small comparing to the detuning between
the cavity and the charge qubit with the values of $198\longrightarrow 202.6%
\text{MHz}$. The appearance of the excited state $|a\rangle $ will
induce noise during the storage process. The magnitude of
coefficient $\beta \sin \theta $ for the excited state decides the
degree of the noise. Just as shown in Fig.~\ref{betat}, the value of
the coefficient is very small with the slowly increase of the
control field and the influence of the noise can be omitted.
 All these results indicate that the charge qubit can be mapped into the
collective excitation of the molecular ensemble in cur- rent
experimental condition effectively without any real photon exchange


\section{Conclusions}

In summary, we have explored and studied a quantum storage scheme for charge
qubit, which is based on the virtual photon exchange between charge qubit
and the collective excitations of polarized molecules. Our investigation
concerns a hybrid system consisting of a superconducting circuit QED system
and an ensemble of polarized molecules. Through the adiabatic elimination
using the Fr\"{o}hlich's transformation, we explicitly derive an effective
coupling between the qubit and the molecular ensemble. It shows that the
effective dark state can be dynamically produced as an entanglement of qubit
plus molecular ensemble, which means that the qubit state can directly be
mapped into the state of molecular ensemble without middle time control. The
corresponding numerical results show the feasibility of this experimental
scheme.

\begin{acknowledgements}
The work is supported by National Natural Science Foundation of
China under Grant Nos. 10547101 and 10604002, the National
Fundamental Research Program of China under Grant No. 2006CB921200.
\end{acknowledgements}


\begin{thebibliography}{Wallraff(2004)}
\bibitem{Knill} K. R. Laflamme and G. J. Milburn, Nature (Londond) \textbf{%
409}, 46 (2001).

\bibitem{Zhou} D. L. Zhou, B. Zeng, Z. Xu, and C. P. Sun, Phys. Rev. A
\textbf{68}, 062303 (2003).

\bibitem{Barenco} A. Barenco, C. H. Bannett, R. Cleve, D. P. DiVincenzo, N.
Margolus, P. Shor, T. Sleator, J. A. Smolin, and H. Weinfurter, Phys. Rev. A
\textbf{52}, 3457 (1995).

\bibitem{Pazy} E. Pazy, I. D'Amico, P. Zanardi, and F. Rossi, Phys. Rev. B
\textbf{64}, 195320 (2001).

\bibitem{Song} Z. Song, P. Zhang, T. Shi, and C. P. Sun, Phys. Rev. B
\textbf{71}, 205314 (2005).

\bibitem{Taylor} J. M. Taylor, C. M. Marcus, and M. D. Lukin, Phys. Rev.
Lett. \textbf{90}, 206803 (2003).

\bibitem{Sunm} C. P. Sun, Y. Li, and X. F. Liu, Phys. Rev. Lett. \textbf{91}%
, 147903 (2003).

\bibitem{Makh} Y. Makhlin, G. Sch\"{o}n, and A. Shnirman, Rev. Mod. Phys.
\textbf{73}, 357 (2001).

\bibitem{charge} Y. Nakamura, Yu. A. Pashkin, and J. S. Tsai, Nature \textbf{%
398}, 786 (1999).

\bibitem{Vion} D. Vion, A. Aassime, A. Cottet, P. Joyez, H. Pothier, C.
Urbina, D. Esteve, and M. H. Devoret, Science \textbf{296}, 886 (2002).

\bibitem{flux} I. Chiorescu, Y. Nakamura, C. J. P. M. Harmans, and J. E.
Mooij, Science \textbf{299}, 1869 (2003).

\bibitem{Mart} J. M. Martinis, S. Nam, J. Aumentado, and C. Urbina, Phys.
Rev. Lett. \textbf{89}, 117901 (2002).

\bibitem{Yu} Y. Yu, S. Han, X. Chu, S. I. Chu, and Z. Wang, Science \textbf{%
296}, 889 (2002).

\bibitem{Pash} Yu. A. Pashkin, T. Yamamoto, O. Astafiev, Y. Nakamura, D. V.
Averin, and J. S. Tsai, Nature (London) \textbf{421}, 823 (2003).

\bibitem{hybi} P. Rabl, D. DeMille, J. M. Doyle, M. D. Lukin, R. J.
Schoelkopf, and P. Zoller, Phys. Rev. Lett. \textbf{97}, 033003 (2006).

\bibitem{holo} K. Tordrup, A. Negretti, and K. M{\o }lmer, Phys. Rev. Lett.
\textbf{101}, 040501 (2008).

\bibitem[Tordrup(2008)]{Tordrup08} K. Tordrup and K. M{\o }lmer, Phys. Rev.
A \textbf{77}, 020301 (2008).

\bibitem[Wallraff(2004)]{Wallraff04} A. Wallraff, D. I. Schuster, A. Blais,
L. Frunzio, R. S. Huang, J. Majer, S. Kumar, S. M. Girvin, and R. J.
Schoelkopf, Nature (London) \textbf{431}, 162 (2004).

\bibitem[Blais(2004)]{Blais04} A. Blais, R. S. Huang, A. Wallraff, S. M.
Girvin, and R. J. Schoelkopf, Phys. Rev. A \textbf{69}, 062320 (2004).

\bibitem[Liu(2001)]{Liu01} Y. X. Liu, C. P. Sun, S. X. Yu, and D. L. Zhou,
Phys. Rev. A \textbf{63}, 023802 (2001); G. R. Jin, P. Zhang, Y. X. Liu, and
C. P. Sun, Phys. Rev. B \textbf{68}, 134301 (2003).

\bibitem{lukin} M. Fleischhauer and M. Lukin, Phys. Rev. A \textbf{65},
022314 (2002).

\bibitem{froli} H. Fr\"{o}hlich, Phys. Rev. \textbf{79}, 845 (1950); Proc.
R. Soc. London, Ser. A \textbf{215}, 291(1952); Adv. Phys. \textbf{3}, 325
(1954).

\bibitem{froli1} S. Nakajima, Adv. Phys. \textbf{4}, 463 (1953).

\bibitem{froli2} H. B. Zhu and C. P. Sun, Chine. Sci. (A) \textbf{30}, 928
(2000); Prog. Chine. Sci. \textbf{10}, 698 (2000).

\bibitem{adi1} C. P. Sun, Phys. Rev. D \textbf{41}, 1318 (1990).

\bibitem{adi2} A. Zee, Phys. Rev. A \textbf{38}, 1 (1988).
\end{thebibliography}
\end{document}